%% file: nu06_maruyama.tex
\input{cuore_def.tex}

\documentclass[letter]{jpconf}
\usepackage{graphicx, epsfig, amssymb}

\begin{document}
\title{Cryogenic Double Beta Decay Experiments: \\
CUORE and CUORICINO}

\author{Reina Maruyama, for the CUORE Collaboration \cite{CUORE_collab}}

\address{Lawrence Berkeley National Laboratory \& University of Wisconsin, Madison, USA}

\ead{rmaruyama@lbl.gov}

\begin{abstract}
Cryogenic bolometers, with their excellent energy resolution, flexibility in material, and availability in high purity, are excellent detectors for the search for neutrinoless double beta decay.  Kilogram-size single crystals of \teod\ are utilized in CUORICINO for an array with a total detector mass of 40.7\,kg.  CUORICINO currently sets the most stringent limit on the halflife of \tect\ of $\mathrm{T}_{1/2}^{0\nu} \geq 2.4\times 10^{24}$\,yr (90\% C.L.), corresponding to a limit on the effective Majorana neutrino mass in the range of $\langle m_{\nu}\rangle \leq$ 0.2\,--\,0.9\,eV.  Based on technology developed for CUORICINO and its predecessors, CUORE is a next-generation experiment designed to probe $\langle m_{\nu}\rangle$ in the range of 10\,--\,100\,meV.  Latest results from CUORICINO and overview of the progress and current status of CUORE are presented.
\end{abstract}

\section{Introduction}

The search for neutrinoless double beta decay (\BBz) has become one of the top priorities in the field of neutrino physics since the discovery of neutrino oscillations in atmospheric\cite{SuperK01}, solar\cite{SNO02}, and reactor\cite{KamLand03} experiments.  An overview and current status of double-beta decay physics and experiments were given in earlier talks by Hirsch, Simkovic, Elliott, and Barabash\cite{DBDTalks1}.  The need to verify the claim of the observation of \BBz\ by a subset of Heidelberg-Moscow germanium experiment\cite{klapdor} has also been presented in these talks.  The most stringent limit on the effective mass of Majorana neutrinos comes from two $^{76}$Ge experiments, Heidelberg-Moscow\cite{HM01} and IGEX\cite{IGEX02}. CUORICINO which is searching for \BBz\ in \tect\ follows closely behind\cite{arnaboldi05}. NEMO-3 is capable of a multiple-isotope search for double-beta decay events, and with its tracking capabilities, has excellent sensitivity to \BBd\cite{NEMO05}.

A number of experiments are currently at various stages of development to probe the degenerate mass hierarchy region of the neutrino mass spectrum and into the inverted hierarchy, many of which are represented at this conference\cite{DBDTalks2}.  CUORE (Cryogenic Underground Observatory for Rare Events) is one such experiment, to be located at the Gran Sasso National Laboratory (LNGS). It will consist of 988 bolometers of \teod\ crystals, with a total mass of 741\,kg.  Because of the high isotopic abundance of 34\%, 204\,kg of \tect\ is available for \BBz\ without isotopic enrichment, making CUORE both timely and significantly less expensive than other experiments.  CUORE's modular design and flexibility will also allow future searches in other isotopes of interest.  It is imperative to carry out double beta decay searches in multiple isotopes, both to improve the nuclear matrix calculations necessary to extract the effective neutrino mass, and to ensure that the observation of a line at the expected energy is not a result of an unidentified background.

Double beta decay experiments can be divided into three categories: indirect measurements such as geochemical analyses, direct measurements with the source being separate from the detector, and direct measurements with a detector that also acts as the source.  Bolometers belong to the last category\cite{Moseley94}. When the source is the same as the detector, the source mass is maximized while materials that could potentially contribute to the background are minimized.  In bolometers, the deposited energy is measured thermally, therefore the entire energy of a decay event is fully accounted for. At low temperatures (the operating temperature for CUORICINO is 8\,mK), the heat capacity of crystals is proportional to the cube of the ratio of the operating and Debye temperatures.  The energy released in a single particle interaction within the crystal is clearly measurable as change in temperature of the entire crystal. The temperature change is measured by neutron transmutation doped (NTD) germanium thermistors which are optimized to operate at these temperatures\cite{haller84, haller95}.  The energy resolution of cryogenic bolometers rivals that of germanium detectors, and 5\,keV FWHM resolution at 2.5\,MeV is readily achievable.  

\section{CUORICINO: Results and Performance}
\begin{figure}[htbp]
\begin{center}
\includegraphics[width=3.5pc]{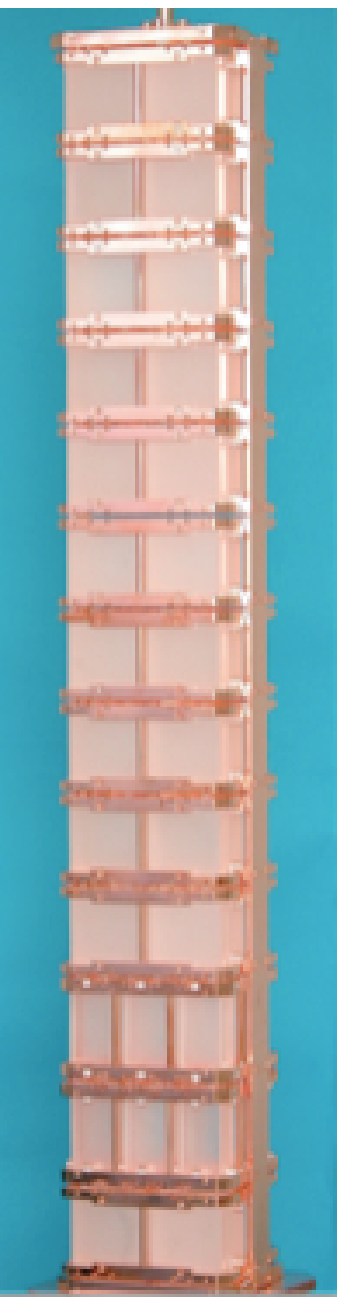}
\includegraphics[width=19pc]{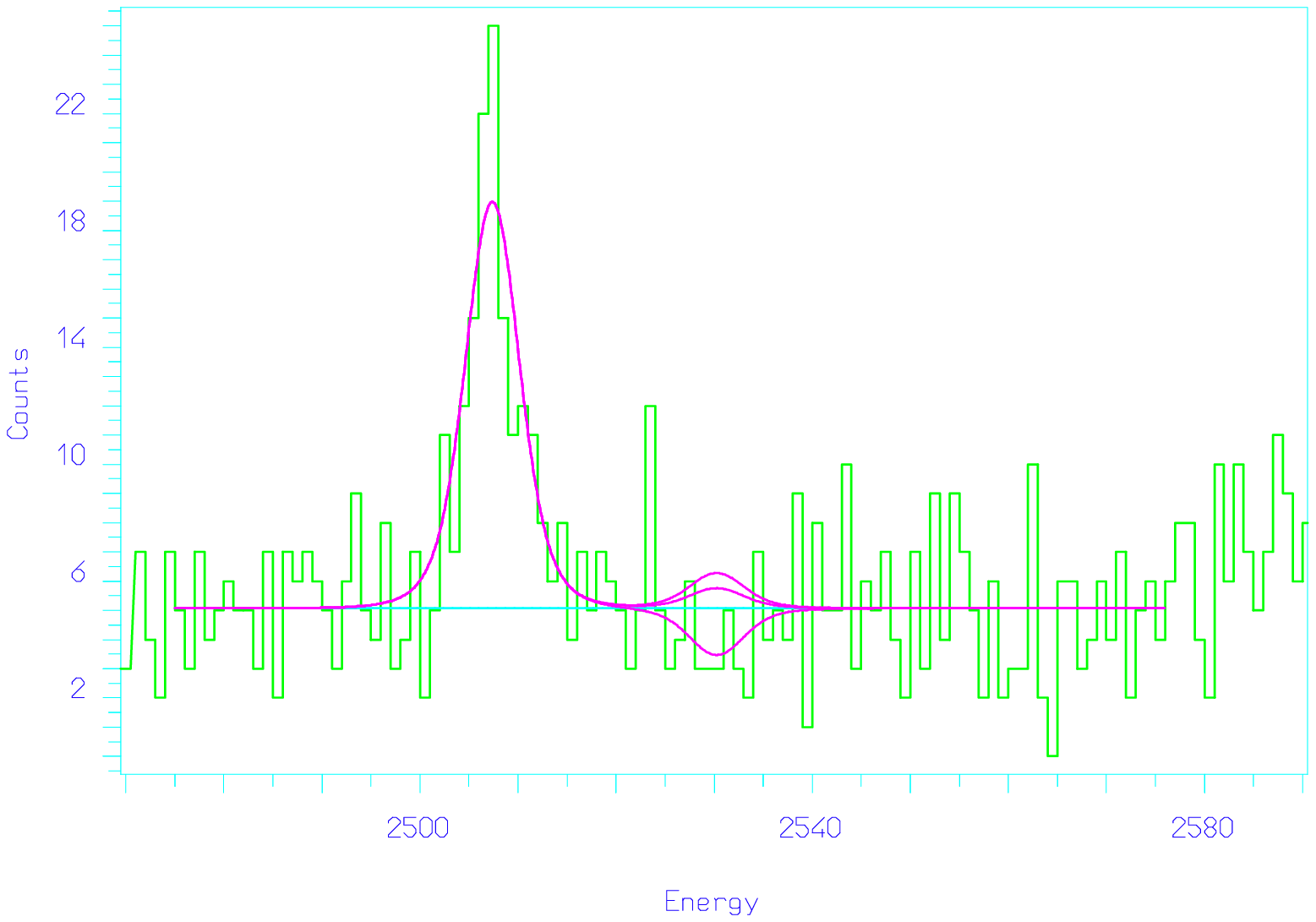}
\caption{\label{Qino}Left: Photo of the CUORICINO tower before Cu thermal shields were installed.  Right: CUORICINO summed background energy spectrum in the \tect\ \BBz\ region.  The peak at 2505\,keV is the sum peak from two \coss\ gamma lines. \BBz\ signal from \tect\ is expected at \Qbb.  No evidence of DBD is seen.}
\end{center}
\end{figure}

CUORICINO started taking data in April 2003 at LNGS and is now producing competitive results with those achieved by the germanium experiments.  It will continue to run until CUORE has been constructed and is ready to take data.  First results from CUORICINO were published recently which included data from a total exposure of 3.09\,kg\dot yr of \tect\cite{arnaboldi05}.  Here we report on an update that includes the data up to May 2006 with a total of 8.38\,kg-yr of \tect\  (see Fig.\,\ref{Qino})\cite{CapelliPoster}.  No evidence for excess counts is observed at \Qbb, the expected Q-value for \BBz\ for \tect.   The absence of any excess events above backgrounds in the region of interest gives a limit of $\mathrm{T}_{1/2}^{0\nu} \geq 2.4\times 10^{24}$\,yr (90\% C.L.) on the \BBz\ decay rate of \tect. This corresponds to an effective neutrino mass of $\langle m_{\nu}\rangle \leq$ 0.18\,--\,0.94\,eV, the range reflecting the spread in QRPA nuclear matrix element calculations (see~\cite{arnaboldi05} for list). The background measured in the \BBz\ region of interest is \QinoBG\,\bkgdunit. 

CUORICINO is roughly one-twentieth the size of CUORE, and much of the technology that will be used in CUORE was used to build CUORICINO.  It consists of 62 \teod\ crystals with a total mass of 40.7\,kg.  The crystals are arranged in a tower, 11 levels each containing four crystals \ciccio\ in size weighing $\sim 790$\,g and 2 levels each housing 9 crystals, \magro\ in size, weighing $\sim 330$\,g (see Fig.\,\ref{Qino}).   All \ciccio\ crystals and all but four of the \magro\ crystals are made from tellurium of natural abundance.  Two of the \magro\ crystals are enriched to 75\% \tect\ and two are enriched to 82.3\% \tecv. The average resolution in the \BBz\ region, measured with the 2615\,keV \tld\ line during calibration runs, is $\sim8$\,keV.  
 In 3 years of running with the present background level, CUORICINO will achieve a half-life sensitivity for $0\nu\beta\beta$ decay of $7.1\times 10^{24}$ yr, corresponding to an effective mass on the order of 300\,meV. 

\section{CUORE}

\begin{figure}[htbp]
\begin{center}
 \includegraphics[width=0.27\textwidth]{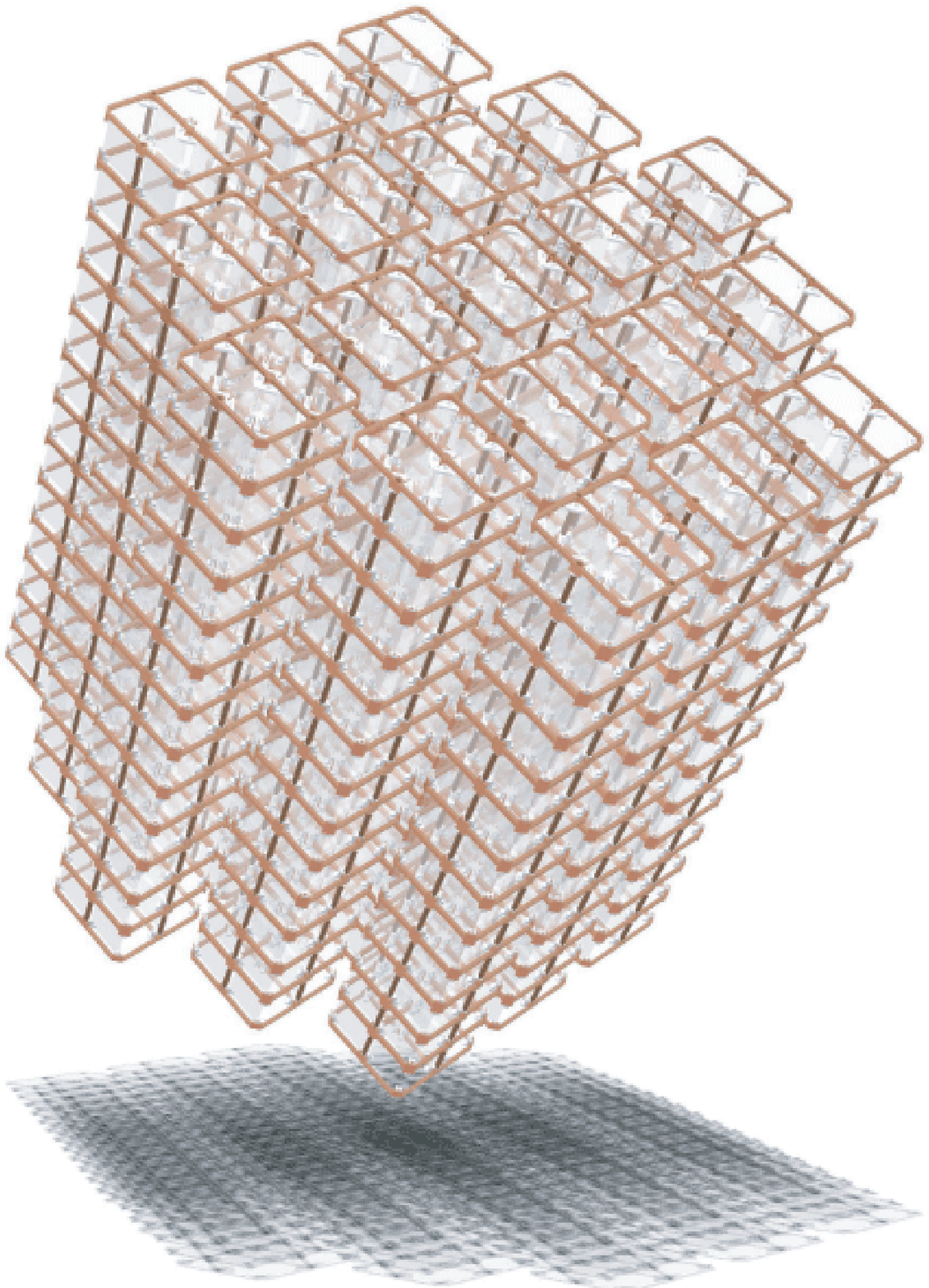}
 \includegraphics[width=0.18\textwidth]{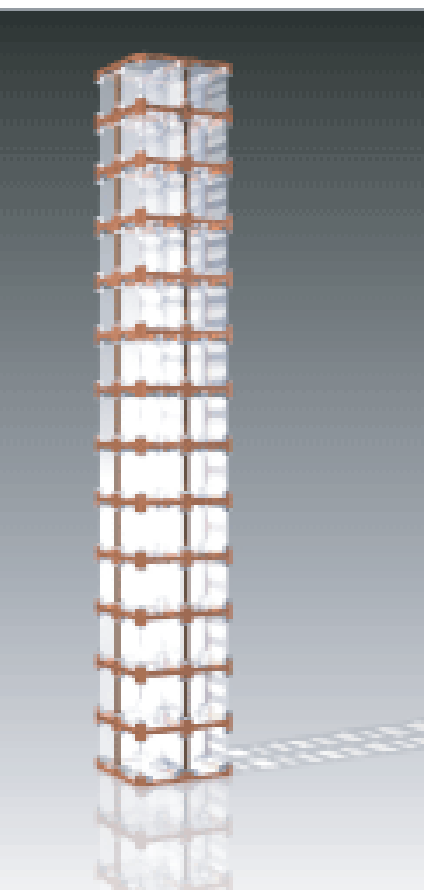}\hspace{2pc}
 \begin{minipage}[b]{18pc}
\caption{ \label{fig:cuore_cyl} Left: The CUORE detector consisting of a close-packed array of 19 towers with a total of 988 crystals. Right: One of the 19 towers of the CUORE detector array, similar to the one operating in the CUORICINO experiment.}
\end{minipage}
\end{center}
\end{figure}

CUORICINO also serves as an excellent test bed and prototype for CUORE. All critical subsystems of the proposed CUORE detector are based on the design of CUORICINO. CUORE will consist of an array of 988, \ciccio\ \teod\  bolometers arranged in 19 CUORICINO-like towers. The total crystal mass of \teod\ will be 741\,kg, with 204\,kg of \tect\ (see Fig.~\ref{fig:cuore_cyl}).  The entire detector will be housed in a single dilution refrigerator at 10\,mK.  

In 5 years of running with a background of 0.01\,\bkgdunit\ and a resolution of 5\,keV, CUORE expects to have a sensitivity to the half-life of \BBz\ of $\mathrm{T}_{1/2}^{0\nu} \sim 2.1\times 10^{26}$\,yr. This corresponds to an effective neutrino mass of $\langle m_{\nu}\rangle \leq$ 19\,--\,100\,meV, with the spread coming from the uncertainty in matrix element calculations.  If we are able to reduce the background to 0.001\,\bkgdunit, the sensitivity will extend to $\mathrm{T}_{1/2}^{0\nu} = 6.5\times 10^{26}$\,yr (11\,--\,57\,meV).  The main technical challenges will be to control the background levels, ensure that the narrow energy resolution achieved with many of the crystals are uniformly implemented in all crystals, and that all crystals are well calibrated for energy.


A combination of CUORICINO background data, measurements from an independent R\&D setup in Hall~C in LNGS, direct counting with germanium detectors on- and off-site, neutron activation analysis, and other techniques are used to characterize materials and components to be used in CUORE.  The results of these measurements as well as other potential sources of radioactive background (e.g.~environmental activity) are used as input for Monte Carlo simulation.  Estimates of the relative contributions of the main background sources in the ROI in CUORICINO is as follows: 10\pom5\% from U/Th contaminations on the \teod\ surfaces, 50\pom20\% from Cu surfaces (both from the crystal support structure and thermal shielding), and 30\pom10\% from the bulk of the Cu shields.

The sources of backgrounds are divided into three main categories: contamination in the bulk, surfaces, and environmental radioactivity.  Because the Q-value for \tect\ \BBz\ decay is higher than most gamma-lines from U and Th, the only tails of known lines that may contribute to the background for CUORE are the \coss\ and \tld\ lines at 2505\,keV and 2615\,keV respectively.  Alpha events with lines at higher energies can contribute if they deposit only a part of their energy in the crystals, therefore surface contaminations on or near the crystals is of particular concern. 

Other components facing the detector (Teflon stand-offs, heaters used for gain stabilization, and gold wires for signal and other electrical controls) were also tested in the R\&D setup in Hall~C by covering crystal surfaces with a large amount of these materials.  The background seen from these materials was found to be negligible. 

\begin{figure}[htbp]
 \begin{center}
 \includegraphics[width=0.75\textwidth]{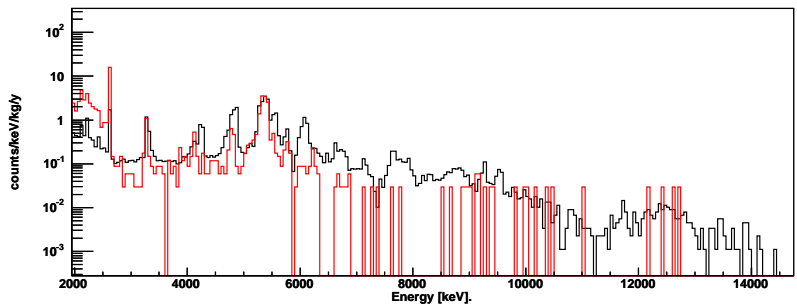}
 \end{center}
 \caption{ \label{fig:cuoricinobackgrounds} CUORICINO background energy spectrum (black) and background spectrum from the R\&D setup in Hall~C (red).}
\end{figure}

Simulations are being refined as more data are being collected with the CUORICINO detector and elsewhere. As of April 2006, we have demonstrated that background reductions of a factor of $\sim 8$ and detector resolutions of 5\,keV are achievable.  Figure~\ref{fig:cuoricinobackgrounds} shows the background spectrum obtained from CUORICINO and the R\&D setup in Hall~C.  The shielding around the Hall~C setup is insufficient to shield much of the $\gamma$'s below 2.6\,MeV, however significant reduction in the $\alpha$ events above 2.6\,MeV is clearly seen.  Effort is underway to further reduce the background by careful material selection and handling procedures. In addition, background rejection through anticoincidence among adjacent crystals will be more effective in the much larger CUORE array and will aid in achieving the background goals.

We have estimated that for the muon flux observed in LNGS (2.5\per 10$^{-8}$ $\mu$/(cm$^{2}\,$s)), muons would produce \ca 0.04 neutrons/day in the polyethylene shield and \ca 25 neutrons/day in the lead shield.  This indicates that neutrons will play a secondary role in the total background compared with other sources of background. In addition, we are planning a series of experiments at the GEANIE facility at Los Alamos National Laboratory to measure cross-sections for neutron-induced reactions on the abundant Te isotopes for neutrons from 1-100 MeV\cite{Nu06_beyond}.  The results of these measurements will then be used in our MC calculations to refine background estimates.  

\section{Beyond CUORE}

The main goal and design of CUORE is to search for \BBz\ decay in \tect.  Its sensitivity can be increased three times ($\sim$60\% improvement in the sensitivity to neutrino mass) by replacing the detectors with enriched crystals. Event identification using multiple signatures from a single event is a powerful tool in reducing backgrounds.  Work is underway to further reduce backgrounds by using Surface Sensitive Bolometers (SSB) and/or scintillating bolometers\cite{Nu06_beyond}.  Every factor of ten reduction in background would increase the halflife sensitivity by a factor of three.

SSB allows us to distinguish surface events and bulk events, especially for $\alpha$-particles.  It consists of the main DBD absorber and thin absorbers attached to the crystal surfaces.  The surface events from either the main crystal absorber or elsewhere would trigger the SSB, and those events could be rejected.  In addition, the additional heat capacity from the thin absorbers alters the pulse shape of the signal from the main absorber\cite{SSB06}. 

Scintillating bolometers would combine heat and scintillation approach already successfully applied in dark matter experiments such as CRESST and ROSEBUD.  Scintillation and heat signals have different sensitivities for nuclear recoils, $\alpha$ particles, and ionizing events 
such as \BBz\ decay.  A CaF$_2$ bolometer has successfully been used\cite{Alessandrello98}, and
the collaboration is currently investigating \teod\ doped with Nb and Mn\cite{Dafinei05}. 

The modular design of the CUORE detector also allows for searches of \BBz\ in other isotopes.  It is possible to create thermal detectors from a variety of materials, and CUORE could investigate \BBz\ in other nuclei.  Several DBD candidates have been tested as thermal detectors: CaF$_{2}$, Ge, MoPbO$_4$, CdWO$_4$, and TeO$_2$. Possible crystals for Nd are under development.  

\section{Conclusion}
Cryogenic bolometers, with their flexibility in material choice and the ability to scale up to the ton-scale are ideal for large-scale detectors for double-beta physics experiments.  CUORE aims to probe the Majorana nature of the neutrino, with a sensitivity to the neutrino mass deep into the inverted mass hierarchy.  CUORICINO is currently running as the most sensitive \BBz\ experiment, and will continue until CUORE comes online. Much of the technology has been tested for CUORE, and a factor of 8 reduction from the radioactive background observed in CUORICINO has been achieved.  CUORE has been approved by the advisory Commissione II of INFN (Italian Institute of Nuclear Physics) and funding has been allocated in 2005.  The CUORE experiment was approved by the Scientific Committee of LNGS in 2004, and preparations of the laboratory space and the construction of CUORE are underway.

\end{document}

%% file: cuore_def.tex
\def\tect{$^{130}$Te}
\def\tecv{$^{128}$Te}

\def\tld{$^{208}$Tl}

\def\coss{$^{60}$Co}

\def\ca{$\sim$}

\def\dot{$\cdot$}
\def\pom{$\pm$}

\def\teod{TeO$_2$}
\def\be{\begin{equation}}
\def\ee{\end{equation}}

\def\per{$\times$}

\def\ciccio{5$\times$5$\times$5\,cm$^3$}
\def\magro{3$\times$3$\times$6\,cm$^3$}

\def \BBz{$0\nu\beta\beta$}
\def \BBd{$2\nu\beta\beta$}

\def \bkgdunit{counts/keV\dot kg\dot yr}

\def \Qbb{2530\,keV}
\def \QinoBG{$ 0.18 \pm 0.01$}